\documentclass[manuscript,screen]{acmart}

\AtBeginDocument{%
  }

\copyrightyear{2026}
\acmYear{2026}
\setcopyright{cc}
\setcctype{by}
\acmConference[CHI '26]{Proceedings of the 2026 CHI Conference on Human Factors in Computing Systems}{April 13--17, 2026}{Barcelona, Spain}
\acmBooktitle{Proceedings of the 2026 CHI Conference on Human Factors in Computing Systems (CHI '26), April 13--17, 2026, Barcelona, Spain}
\acmDOI{10.1145/3772318.3791869}
\acmISBN{979-8-4007-2278-3/2026/04}

\begin{document}

\title[Deaf/HH Access to Intelligent Personal Assistants: Voice vs LLM Touch]{Deaf and Hard of Hearing Access to Intelligent Personal Assistants: Comparison of Voice-Based Options with an LLM-Powered Touch Interface}

\author{Paige S. DeVries}
\email{paige.devries@gallaudet.edu}
\orcid{0009-0009-9815-7121}
\affiliation{%
  \institution{Gallaudet University}
  \city{Washington}
  \state{D.C.}
  \country{USA}
}

\author{Michaela Okosi}
\email{michaela.okosi@gallaudet.edu}
\orcid{0009-0006-5224-6696}
\affiliation{%
  \institution{Gallaudet University}
  \city{Washington}
  \state{D.C.}
  \country{USA}
}

\author{Ming Li}
\email{ming.li@gallaudet.edu}
\orcid{0009-0009-0177-5178}
\affiliation{%
  \institution{Gallaudet University}
  \city{Washington}
  \state{D.C.}
  \country{USA}
}

\author{Nora Dunphy}
\email{noradunphy@berkeley.edu}
\orcid{0009-0004-5011-5188}
\affiliation{%
  \institution{University of California, Berkeley}
  \city{Berkeley}
  \state{CA}
  \country{USA}
}

\author{Gidey Gezae}
\email{gjg5425@psu.edu}
\orcid{0000-0001-5566-7951}
\affiliation{%
  \institution{Pennsylvania State University}
  \city{State College}
  \state{PA}
  \country{USA}
}

\author{Dante Conway}
\email{dante.conway@gallaudet.edu}
\orcid{0009-0000-7297-446X}
\affiliation{%
  \institution{Gallaudet University}
  \city{Washington}
  \state{D.C.}
  \country{USA}
}

\author{Abraham Glasser}
\email{abraham.glasser@gallaudet.edu}
\orcid{0000-0003-1763-4352}
\affiliation{%
  \institution{Gallaudet University}
  \city{Washington}
  \state{D.C.}
  \country{USA}
}

\author{Raja Kushalnagar}
\email{raja.kushalnagar@gallaudet.edu}
\orcid{0000-0002-0493-413X}
\affiliation{%
  \institution{Gallaudet University}
  \city{Washington}
  \state{D.C.}
  \country{USA}
}

\author{Christian Vogler}
\email{christian.vogler@gallaudet.edu}
\orcid{0000-0003-2590-6880}
\affiliation{%
  \institution{Gallaudet University}
  \city{Washington}
  \state{D.C.}
  \country{USA}
}

\renewcommand{\shortauthors}{DeVries et al.}

\begin{abstract}
We investigate intelligent personal assistants (IPAs) accessibility for deaf and hard of hearing (DHH) people who can use their voice in everyday communication. The inability of IPAs to understand diverse accents including deaf speech renders them largely inaccessible to non-signing and speaking DHH individuals. Using an Echo Show, we compare the usability of natural language input via spoken English; with Alexa's automatic speech recognition and a Wizard-of-Oz setting with a trained facilitator re-speaking commands against that of a large language model (LLM)-assisted touch interface in a mixed-methods study. The touch method was navigated through an LLM-powered `task prompter,' which integrated the user's history and smart environment to suggest contextually-appropriate commands. Quantitative results showed no significant differences across both spoken English conditions vs LLM-assisted touch. Qualitative results showed variability in opinions on the usability of each method. Ultimately, it will be necessary to have robust deaf-accented speech recognized natively by IPAs.
\end{abstract}




\maketitle

\section{Introduction}

Intelligent personal assistants (IPAs), such as Amazon Alexa~\cite{lopatovska2019talk} and  Google Assistant~\cite{Google_nodate}, offer the convenience of voice-controlled functionality but present significant accessibility challenges for Deaf and Hard of Hearing (DHH) individuals ~\cite{ballati2018hey,efthimiou2020proceedings,glasser2017deaf,glasser2019automatic,pradhan2018accessibility} and even more limited options for DHH non-signers with dysarthric speech. Currently, Automatic Speech Recognition (ASR) still struggles to understand the variability of deaf speech~\cite{rodolitz2019accessibility,bigham2017deaf,jaddoh2023interaction,qian2023survey,liu2021recent}. 

The level of dysarthric speech often depends on when the individual lost their ability to hear and the level of hearing loss. The consensus on a functionally equivalent input method for DHH users still remains a question~\cite{kafle2020artificial,rodolitz2019accessibility}. Current IPAs restrict non-typical verbal input options due to their over-reliance on ASR technologies unable to adapt to non-standard speech patterns. Non-verbal approaches currently pose a significant latency that negatively impacts user experience and accessibility~\cite{fok2018towards}. As the development and daily integration of these technologies grows exponentially, the need for interfaces designed for DHH usability becomes more urgent.


There has been much investigation of sign language input as a possibility in lieu of speech interfaces, but very little research on those who are DHH and use their voice. With respect to DHH people who speak, this paper aims to provide much-needed research into the the usability of input methods for IPAs that are technically feasible today. This group of DHH people currently, in principle, has three options for interacting with current-generation IPAs: (1) their own speech, (2) using specially trained software for their deaf accent, which re-speaks the recognized commands into a form that IPAs can understand, and (3) touchscreen interfaces.

With respect to (1), as mentioned above, many DHH people express dysarthric speech that makes it difficult for them to interact with IPAs directly. Nevertheless, it is important to study to what extent this is a problem in current practice. With respect to (2), re-speaking software for dysarthric speech is commercially available~\cite{voiceittVoiceittInclusive}, but induces a delay and requires the user to train the system in advance. With respect to (3), although touchscreen input may seem like a sufficient alternative, its current usability is less efficient than that of natural language input, as previously shown in studies on DHH participants~\cite{tran2024assessment} in part due to the time required to navigate the user interface and type commands. 

However, with touchscreens, historically, such approaches have not been context-aware enough to provide the most relevant options that a user might select at any given time. With the commercialization of generative artificial intelligence (AI) that utilizes large language models (LLMs), such as ChatGPT, there are new opportunities for efficient and adaptable user interfaces through the use of situational context and previous user commands. This opens up new opportunities for improvements to touch-based interaction that can benefit DHH people. 

\begin{figure}[h]
  \centering
  \includegraphics[width=3.2in,alt={A user is selecting options generated by the Large Language Model via a tablet showing a selection of 8 buttons.}]{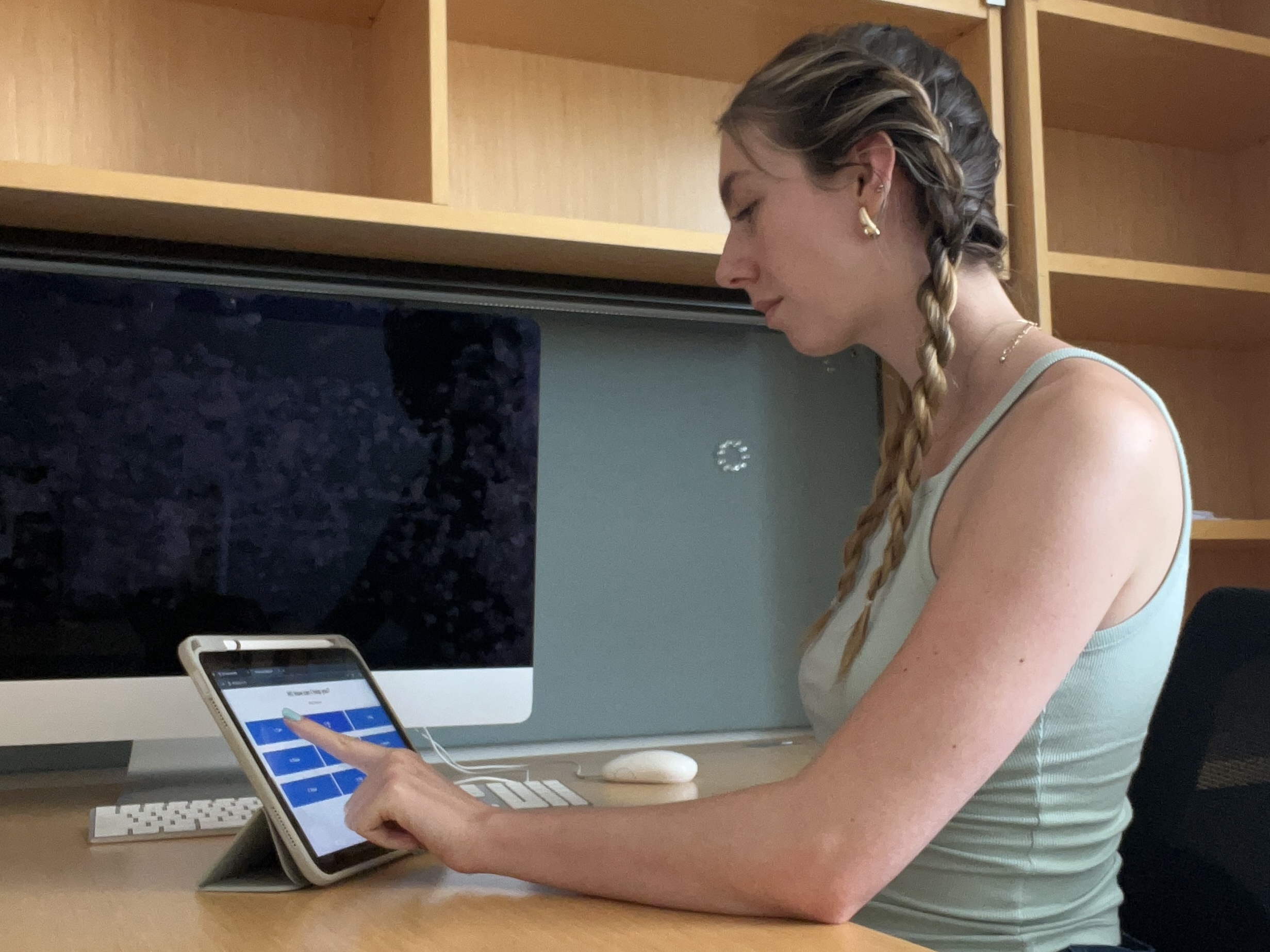}
  \caption{A user selecting LLM-populated options on a touchscreen in our LLM Touch interface.}
  \label{fig:inputmethods}
  \Description{A user is selecting options generated by the Large Language Model via a tablet showing a selection of 8 buttons.}
\end{figure}

In this paper, we aim to assess and compare the usability of what currently appear to be the most promising options for non-signing DHH people to interact with IPAs: (1) a novel LLM-assisted touch interface, shown in Figure~\ref{fig:inputmethods}, which leverages the situational context and ongoing interaction with each user in an attempt to make touch-based options more effective, and (2) spoken English usability, divided into two approaches: (2a) natural deaf speech with the built-in ASR technology in IPAs, and (2b) re-speaking deaf-accented dysarthric speech to an IPA in more easily intelligible English.

\subsection{Study Approach}
\label{sec:approach}
The approach chosen for this paper was a mixed method study that allowed participants to explore three input methods (LLM-assisted touch, natural deaf speech, and re-speaking). For re-speaking, we originally considered using automatic speech recognition technologies that have been developed for those that have non-standard speech patterns (like VoiceITT and Project Relate~\cite{voiceittVoiceittInclusive,researchProjectRelate}), but ultimately decided against it as it would have required participants to spend hours on training the system. From an experimental design standpoint, this would have posed a risk of the training experience unduly influencing the participant ratings compared to the other input methods that we tested. 

To work around this re-speaking barrier, for a valid comparison between speech and touch, we asked participants to test both their natural speech with Alexa's built-in ASR, and Wizard-of-Oz-facilitated re-speaking, where a trained researcher listened to and watched the participant on video, then re-voiced their spoken commands. A human facilitator, in a sense, represents the best case for re-speaking, with no prior training for participants required, and a higher ceiling of understanding deaf-accented speech than the currently available commercial options. We evaluated usability both quantitatively through usability questionnaires and qualitatively through user interviews to gain further insight into the usability scores.

\subsection{Research Questions}
\label{sec:rq}
The research questions for the study presented in this paper are:

\begin{itemize}
\item RQ1: How do DHH IPA users who use their voice perceive the usability and usefulness of LLM-generated task suggestions compared to deaf speech input?
\item RQ2: How do the satisfaction ratings of DHH users, who use their voice, compare between natural deaf speech and Wizard-of-Oz Facilitated English speech methods?
\end{itemize}

In the remainder of this paper, we discuss related work, then describe the mixed-method experimental design, provide quantitative and qualitative results, discuss their implications, and note the limitations of the study design.

\section{Related Work}

The limitations of current IPAs demonstrate a need for improved accessibility that AI and machine learning have shown potential in aiding.  The most relevant research contributions thus far include the validation of  deaf accented spoken English as a viable IPA input method, as well as the enhancements LLMs can provide to such interactions. 

\subsection{DHH that use Speech Experiences}

By studying current technology, we can work toward developing smart devices that better assist deaf and hard of hearing individuals. Automatic Speech Recognition (ASR) is becoming increasingly accurate, with Google’s ASR reaching a 5\% word error rate (WER) in 2016-17~\cite{bigham2017deaf,glasser2017deaf,10.1109/ICASSP.2018.8461870}. The increasing supply and demand of ASR-reliant aural interfaces in home environments has exacerbated the gap in modern technology accessible to DHH users.

Several studies have already addressed the need for IPAs with improved accessibility~\cite{turk2014multimodal,docio2020lse_uvigo,mande2021deaf}. One article examined whether ASR software was able to effectively understand Deaf speech. DHH participants who use their voice expressed difficulty while interacting with this technology, which Glasser attributes to variance in their speech patterns~\cite{glasser2017deaf}. 


One particular challenge DHH users who use their voice face when using IPAs is a lack of visual feedback. Blair et al~\cite{blair2020didn}~found that although some users can have successful interactions with these devices, they face challenges relating to output voice quality, background noise, and visual information. To remedy these difficulties, the paper proposes a pitch customization feature to adjust for specific hearing needs (such as hearing aid compatibility) and increased visual feedback (such as the addition of flashing light notifications). Utsuki et al suggest using IPAs to assist deaf and hard of hearing dog owners in identifying potentially dangerous situations. The paper discusses an app to detect when a dog barks and notify users through their smartphone, replacing biological auditory feedback with synthesized visual alerts. Deaf participants were tasked with identifying hazardous situations at home and reported satisfaction with the system's accuracy of recognition, intensity, and light activation ~\cite{utsuki2022towards}. 

The most closely related work to this experiment is described in two papers, which examine various input methods for DHH users interacting with IPAs~\cite{tran2024assessment,devries2024sign}. Exploring different input methods for DHH users when interacting with IPAs is crucial in determining usability and accessibility. Tran et al evaluates the usability of American Sign Language (ASL) input, Tap to Alexa (which is the built-in Alexa feature that allows users to interact with the IPA using touchscreen tiles and typing in requests with the touchscreen keyboard), and smart home apps (the Alexa app, Fire TV app, and Philips Hue lights app) in a limited-domain smart home environment through the Wizard-of-Oz method. Among the 23 deaf participants, ASL was preferred with a mean SUS of 71.6, followed by Tap to Alexa (61.4), and Apps (56.3). The paper also identified 117 essential signs for IPA system understanding, discussed linguistic phenomena that need to be addressed for ASL recognition, and suggested that wake gestures simulate attention-grabbing methods commonly used in Deaf culture \cite{tran2024assessment}. Devries et al evaluates the usability of different types of IPA input for Deaf users in a kitchen setting. After participants got their hands dirty, they interacted with Amazon Alexa through ASL and touch-based input. Post-experiment survey results show that participants strongly preferred ASL input and 85\% expressed high interest in an ASL-recognizing IPAs. The low usability rating of smart home apps promotes the need for research into a more dynamic touch-based system ~\cite{devries2024sign}. 

The inaccessibility of aural output has partially been addressed by supplemental captions, though these devices still require spoken input. Current ASR technology struggles to recognize dysarthric speech accents, including that of people with motor disability, and Deaf and Hard of Hearing people~\cite{jaddoh2023interaction,qian2023survey,liu2021recent}. If this continues, the only preexisting solution requires the use of external text-to-speech (TTS) applications. However, TTS does not serve as a functional equivalent to speech as typing remains 3-4 times slower than speaking~\cite{bigham2017deaf}. Additionally, TTS applications do not account for intonation and vary from a natural cadence, lessening the accuracy of aurally-reliant devices. Even when implemented successfully, the output of strictly audio-based devices remain inaccessible.

\subsection{Large Language Models}
Integrating LLMs with IPAs has the potential to enhance accessible interactions for DHH users who use their voice. Current IPAs have the ability to execute preset spoken commands but lack a broader contextual understanding of their interactions with their users. Existing models attempt to identify user intent through a rule-based, keyword recognition system and fail to comprehend multi-turn conversations. In the areas where IPAs fall short, LLMs excel. LLMs bridge the gap between computers and human language by understanding natural language input and generating written text that is both coherent and contextually aware. Combining the individual capabilities of these two technologies has immense potential for the accessibility, usability, and complexity of interactions with personal assistants. Several studies have investigated the use of LLM-powered IPAs for speech input~\cite{mahmood2023llm,tran2024assessment,vu2024gptvoicetasker,yoshitake2023amazon}. 

An added benefit of integrating LLMs into IPAs includes the opportunity to allow users to “access real-time data, enjoy immersive generative AI-enhanced games, perform tasks like booking restaurant reservations, getting a succinct summary of a trending news story, and more”~\cite{yoshitake2023amazon}.  Previous studies have begun to explore the integration of IPAs and LLMs, but the body of work remains limited. This study seeks to expand existing literature on the combined technologies by incorporating the perspectives of both DHH researchers and users. 


The interaction patterns demonstrated by LLM-assisted IPAs support richer interactions across tasks and succeed in most scenarios that rely on recognition of intent ~\cite{mahmood2023llm}. LLMs have also demonstrated their effectiveness as valuable enhancements to virtual assistants for mobile devices. GPTVoiceTasker uses LLMs to transform voice commands into automation of typical smartphone interactions, such as scrolling, tapping, and inputting text. The smartphone IPA understands commands and executes tasks while collecting user data to predict future users input. This technology was received positively by participants and was able to accelerate task completion by over 30\%~\cite{vu2024gptvoicetasker}. 


The power of LLM integration extends to more mainstream research areas outside of the scope of accessibility. One paper proposes a framework for using LLMs to create more dynamic user interfaces (UI)~\cite{wasti2024large}. By identifying user needs and predicting their actions, the model acts as a dynamic intermediary between the user and UI. In their approach, users interact with software by voicing their needs in natural language instead of using conventional buttons and menus. This framework demonstrates the potential to revolutionize the way people interact with software by integrating LLMs into user interfaces. Zhao et al incorporates LLMs to improve recommendation systems~\cite{zhao2024let}. Their framework, ToolRec, guides the recommendation process more precisely by employing LLMs to act as "surrogate users". ToolRec proved to be more effective and accurate than conventional techniques, showing promise for LLMs' role in creating more personalized recommendation systems.

\section{Methods}

Each participant interacted with the Alexa device using three input methods: Natural Deaf Speech, Wizard-of-Oz Facilitated English speech (as the stand-in for commercial re-speaking software, see also Section~\ref{sec:approach}), and LLM-assisted touch input.  

\subsection{Recruitment and Participant Demographics}

We recruited 20 participants through outreach from a university and communities near the area who met the following criteria: (1) 18 years of age or older, (2) identify as Deaf, Hard of Hearing, or having hearing loss, and (3) use spoken English in everyday communication (or are interested in using speech with voice assistants). All procedures were approved through the required ethics reviews. Participants were additionally informed that their command selections were transmitted to ChatGPT-4o for processing, with all personally identifiable information removed. Each participant was paid for their time. 

We had 7 male participants (35\%), 1 non-binary participant (5\%), 1 who preferred not to disclose (5\%), and 11 female participants (55\%). 10 participants identified as Deaf (50\%), 2 were Deaf-Blind (10\%), and 8 were Hard of Hearing (40\%). Participants' ages ranged from 20 to 75 years old, with an average of 33 and a median of 26 years. 8 participants reported having cochlear implants (40\%), 11 reported using a hearing aid (55\%), and one reported not using any hearing devices (5\%).  3 participants identified as Asian (15\%), 6 participants identified as Black (30\%), 1 identified as Hispanic, and the remaining 10 identified as White (50\%).

We asked participants their preferred communication languages and 15\% listed ASL, 5\% listed ASL and Pidgin Signed English (PSE), 10\% listed ASL, PSE, spoken English, and written English, 5\% selected ASL, PSE, and written English, 15\% selected ASL and spoken English, 35\% listed ASL, spoken English, and written English, 10\% listed ASL and written English, and 5\% listed spoken English.

9 participants reported having a voice assistant at home (45\%), and the other 11 did not (55\%). One reported they use their device frequently, two to three times a week (11\%). Two participants reported occasional use, once per week (22\%). Two others reported rare usage, around once every two weeks (22\%).  Three participants reported using their at-home voice assistant very rarely, at once per month (33\%). Lastly, one participant reported that they never interact with the voice assistant present in their home (11\%). Of the eight participants that reported interacting with the voice assistant in their home, five reported using them for setting timers/reminders/alarms, five reported using them for playing music, four reported using them for news and forecasts, one reported finding movies to watch, one reported controlling [other] smart home devices, one reported finding recipes for cooking, and one reported also using them for unspecified other reasons. 

Three participants have a device with a touchscreen (33\%), while six of these participants have a device without one (66\%). Of the three participants whose devices have a touchscreen, one reported occasional use at once per week, and the other two reported rare use, once every two weeks. Seven participants reported using their voice with the device in their home (77\%), while the other two do not (22\%). Of the seven participants who use their voice with their at-home voice assistant device, one reported being understood by the device all of the time, two reported being understood more than half of the time, two reported being understood half of the time, and the last two reported being understood less than half of the time.

\subsection{Materials}

The study was conducted in a simulated smart home environment using an Amazon Echo Show device with captioned responses, as well as two Philips Hue smart lights that could be controlled via Alexa. The user interface was run on a Fire HD 8 (12th generation) tablet for touch input, the details of which are described in Section~\ref{sec:llmarch}. To support the Wizard-of-Oz Facilitated English condition with re-speaking, a webcam was connected to a MacBook Air to provide a remote video feed to the Facilitated English researcher, and a Bluetooth speaker was used to issue the researcher's spoken English commands to Alexa, emulating the setup described in prior published work~\cite{tran2024assessment}.

\subsection{Procedures}
\label{sec:procedures}

As mentioned above, each participant interacted with the Alexa device using three input methods: Natural Deaf Speech, Wizard-of-Oz Facilitated English speech (see also Section~\ref{sec:approach}), and LLM-assisted touch input. 

The study was conducted in person and took up to 90 minutes per session. Participants were guided throughout the session in their preferred language—ASL or spoken English—by one of two DHH researchers, one fluent in ASL and the other in spoken English. One of the researchers stood by to provide technical support if needed, while the remote researcher (“the Wizard”) supporting the Facilitated English condition was in a separate room. To avoid any confusion for participants who may hear the researcher's voice due to wearing cochlear implants or hearing aids, we explained that they should disregard any interference in the condition and informed them that we would provide more details during the interview portion of the study.

After providing consent and completing a pre-experiment survey asking about demographic information and prior experience with IPAs, participants experienced all three input methods. Before each input method, participants completed a short exploration period (approximately 5 minutes, 2–3 practice commands). They then completed a list of 10 predefined tasks, which took roughly 10 minutes per condition (see Appendix \ref{app:tasklist}). Task content was consistent across conditions with small variations (e.g., light color changes). The order of conditions was counterbalanced.

For the Natural Deaf Speech condition, participants spoke directly to the Echo Show using any preferred wake word. If the device did not recognize their speech, they could repeat the command as many times as they wished or choose to skip the condition.

For the Wizard-of-Oz Facilitated English condition, participants spoke toward a webcam placed above the Echo Show, similar to the setup in prior work on ASL and IPA compatibility~\cite{glasser2022analyzing,tran2024assessment,rodolitz2019accessibility,wojtanowski2020alexa}. The Wizard viewed the video feed and re-voiced participants’ utterances to Alexa using the wake word “Ziggy” to avoid wake word conflicts.

For the touch condition, participants used a Fire HD tablet to navigate LLM-generated option based on Figure \ref{fig:inputmethods}. The Fire HD tablet used text-to-speech (TTS) to voice the entered commands for the Echo Show. During our testing, we noticed that the Alexa device did not respond to a robotic TTS voice using the wake word, so we recorded a human voice to add to preface each TTS command given. Participants could also type their command with an on-screen keyboard; the UI pre-filled the partial command based on prior selections.

After each condition, participants completed usability questionnaires. After all tasks were completed, participants filled out a post-experiment survey and participated in a semi-structured interview.

\subsection{Measures and Data Collection}
\label{sec:measures}


The System Usability Scale (SUS) is commonly used in HCI contexts to determine how usable a system is. We chose it as a measure for two additional reasons: First, we could give participants the choice to use the written English SUS questionnaire or a psychometrically validated ASL translation (ASL-SUS)~\cite{berke2019design}. This is important because some of the English SUS wording can be confusing for DHH, who may not be native English speakers. Second, these measures make our quantitative findings comparable to past related works, especially those of Tran et al and DeVries et al~\cite{tran2024assessment,devries2024sign}. In addition to SUS, we also provided the Adjective Scale Score~\cite{bangor2009determining} and Net Promoter Score (NPS)~\cite{sauro2012predicting}. Like SUS, they also have validated ASL versions and have been used in past comparable work. The adjective scale is directly correlated with the SUS, but provides a simpler question that assesses usability holistically. The NPS asks how likely it is that a user will recommend a product to someone else. The SUS usually --- but not always --- predicts NPS scores, and provides a proxy measure for how excited participants are about each respective input method. These three metrics were applied after every condition, resulting in a total of nine measurements for each participant in the full study.

We additionally calculated the Word Error Rate (WER) of each participant's Natural Deaf Speech condition. The trained wizards went through both the video recordings of the participants and the Alexa history in the associated Amazon accounts to transcribe what the participants had spoken, and to compare this against what Alexa had recognized for each respective interaction. Because Alexa tends to ignore filler words such as ``um'' or ``uh,'' and false starts, like ``Alexa, show--- tell me a joke,'' we removed these occurrences from the participants' transcripts. We then used SCLite~\cite{nistTools} to calculate the WER for each participant.

Upon completion of all three conditions and their corresponding usability questionnaires, participants completed a post-experiment survey asking them about their in-experiment experience. Participants then went through a semi-structured interview, which lasted 15 to 30 minutes and was conducted one-on-one in their preferred language. Within the semi-structured interview, we asked participants questions in several categories to gain a deeper insight into their experiences and responses to the usability questionnaires. The categories we discussed in the interviews were: (1) what works well with the speech input methods and what does not (addressing RQ2), (2) what works well with the touch interface and what does not (addressing RQ1), (3) how important hands-free interaction is and what methods they would like to have (addressing both RQ1 and RQ2 and broader implications), and (4) if there was anything else they wanted to share. The full guide can be found in Appendix~\ref{app:guide}.

ASL interviews were recorded and translated into English voice-over by the same researcher who served as the Wizard. Transcripts were generated and checked for accuracy by team members fluent in ASL and English. Interview data were analyzed using Braun and Clarke’s thematic analysis framework~\cite{braun2006using}. Two ASL–English bilingual coders familiar with the study independently coded all units. Discrepancies were resolved through consensus. 

\subsection{LLM Touch System Architecture}
\label{sec:llmarch}

To build the LLM backend infrastructure for generating Alexa commands for the touch condition, we used Python and OpenAI’s GPT-4o Application Programming Interface (API), which were controlled via a cloud-based web server implemented in the Flask framework. The front end, which communicated with the cloud server, consisted of a web application running on a browser on the Fire tablet and relied on HTML5, CSS, and JavaScript. The application was responsible for sending the final commands to Alexa via TTS over the tablet's speakers. Because Alexa did not respond to a TTS-generated "Alexa" wake word, we played a human-recorded 'Alexa', followed by the command phrase through TTS. 

\begin{figure*}[h]
  \centering
  \includegraphics[,alt={An overview of the system shows how the user interacts with the system via a Fire Tablet using touch input. The interface runs a combined HTML 5, CSS, and Java Script UI that presents a list of eight options generated by the LLM. After the user selects an option, the system sends the selection via an asynchronous HTTP request to the Flask server, which communicates with the GPT API to generate context-sensitive new options or subsequent responses. The selection process is continuously updated, and the finalized command is read aloud by the TTS module, enabling interaction with Alexa.}]{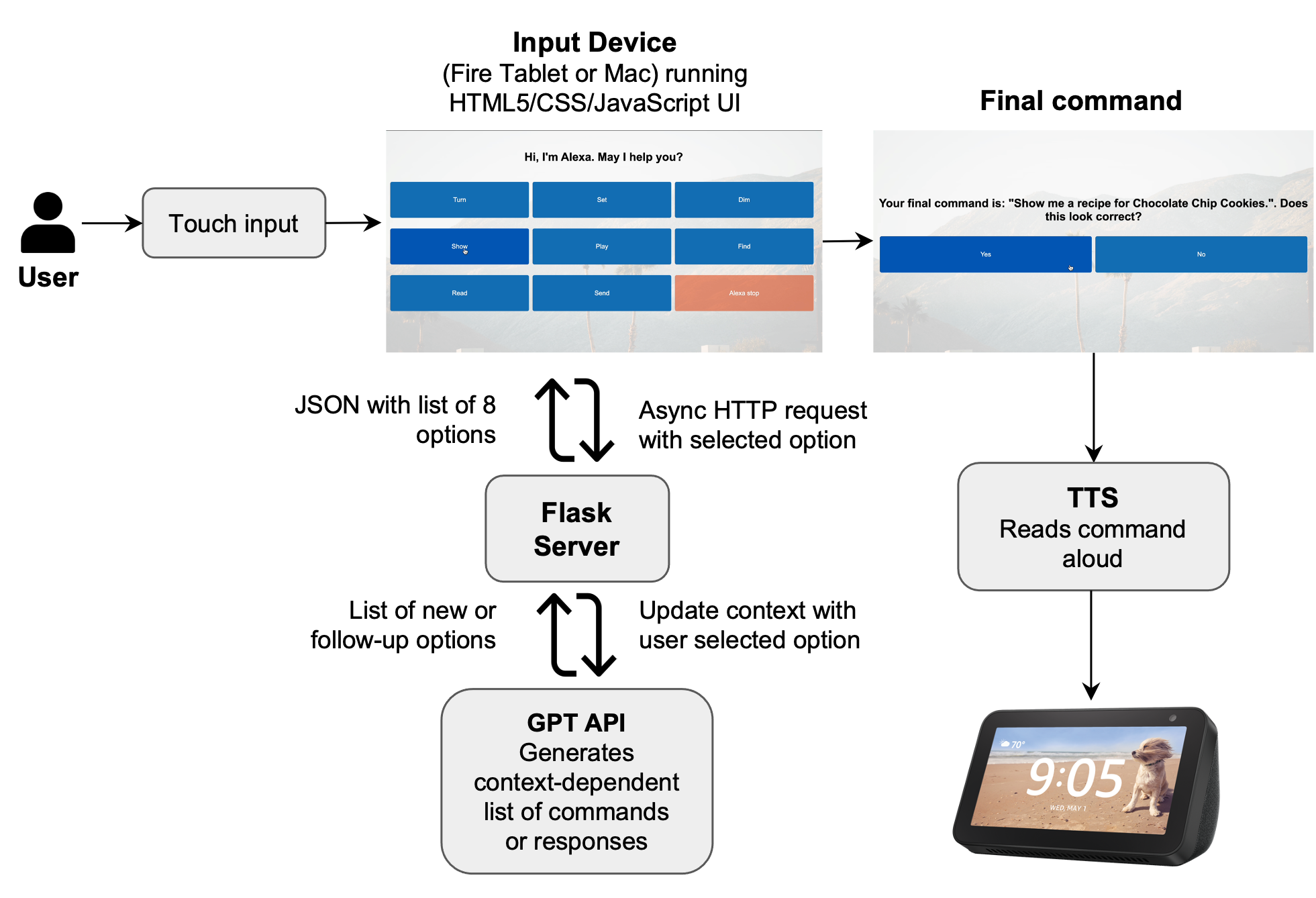}
  \caption{System overview. The user interacts with a tablet which uses Web technologies to communicate with an LLM backend to iteratively generate the Alexa command.}
  \label{fig:system}
  \Description{An overview of the system shows how the user interacts with the system via a Fire Tablet using touch input. The interface runs a combined HTML 5, CSS, and Java Script UI that presents a list of eight options generated by the LLM. After the user selects an option, the system sends the selection via an asynchronous HTTP request to the Flask server, which communicates with the GPT API to generate context-sensitive new options or subsequent responses. The selection process is continuously updated, and the finalized command is read aloud by the TTS module, enabling interaction with Alexa.}
\end{figure*}

The overall system is shown in Figure~\ref{fig:system}. The following sections provide a detailed explanation of each component. The code implementing the LLM Touch interface can be accessed at \url{https://github.com/Gallaudet-University/chi2026-ipa}.

\subsubsection{Large Language Model Implementation}
\label{sec:llm}

We used ChatGPT-4o to generate relevant and contextually aware IPA task suggestions. These interactions with the API were determined by prompts included in the Appendix~\ref{app:llmprompts}. To manage ChatGPT’s inherent variability, we fed the API as many context clues and examples (prompts) as possible while allowing freedom in generating the resulting output. We also adjusted the temperature of ChatGPT-4o from 0.7 to 0.2 to decrease randomness of the responses and improve consistency. However, we did not identify a way to keep the ordering of the options completely consistent from command to command, a negative UI attribute that was noted by multiple users in the semi-structured interviews.

Using those prompts and the user's command history, the LLM generated starter verbs for common IPA tasks (for example, “Turn,” “Set,” and “Play”) for the user to select. Based on the selected verb, ChatGPT then auto-completed the phrase with relevant Alexa commands. For example, if the user selected “Set,” the LLM might propose “Set a timer,” “Set an alarm,” and “Set the lights to a color.” 

The LLM code maintained context for the prompts in three global variables: user history, a task list, and a description of the smart environment. The smart environment described the IPA-connected technology in the given room, such as smart lights, smart plugs, and Fire TVs. The program began by querying the API for action verbs based on these global variables. The user selected an option, which was then passed in as an argument to a function that requested the API for a list of tasks beginning with the selected verb. The user’s next selection then ran through a function asking the API if additional details (such as a time for setting an alarm or a color for setting the light shade) were necessary. 

The API’s response to the additional details question determined the code progression from here. If the API returned “No” when asked if additional information was needed, the task selection process was considered complete, and a final prompt generated the full command. If the API returned “Yes,” the task was passed to a different prompt asking the API to generate a question for the required additional information. The API was further prompted to generate suggested answers for the question it returned. Both the question and the list of suggested answers were returned to the front-end user interface. Once the user selected an answer, the task, question, and question answer were passed to a final prompt to generate the full command. In both scenarios, the completed full command was presented to the user for confirmation before being sent to the Echo Show through TTS.

\subsubsection{Flask Server Backend}
\label{sec:flask}

The server back-end component was necessary to allow the front-end user interface to communicate with the ChatGPT-4o API. It served as an intermediary between the user interface running in a web browser, handling secure requests encapsulating user actions, translating these requests into natural language prompts for the LLM, sending them to the ChatGPT API, parsing the LLM responses, and sending the responses back to the user interface. The back-end was implemented on top of the Python-based Flask framework running in Google Cloud. The total round-trip time from user action to response was approximately two seconds. 

\subsubsection{Web App-Based User Interface}

The UI sent user choices to the server back-end and retrieved user options of Alexa commands in response. Each state of the command generation process presented options available to the user as buttons in a 3x3 grid as shown in Figure~\ref{fig:system}. The first eight buttons corresponded to the eight options provided by the LLM (Section~\ref{sec:llm}), while the ninth button enabled the user to issue an "Alexa stop" command at any time to interrupt media playback, timers, and other actions. During the touch condition, participants could type in their command rather than using the buttons on the UI if they preferred. If the keyboard option was chosen, the partial Alexa command based on the buttons selected to date was pre-populated as the keyboard input and made visible to the participant to cut down on time needed to type.
\section{Results}

In the following sections, we describe both quantitative and qualitative results for the experiment. These results begin to address the research questions, which are further interpreted in the discussion.

\subsection{Quantitative Results}

The key quantitative metrics used to answer our research questions were the SUS, NPS, Adjective Scale, WER for the Natural Deaf English condition, and post-experiment survey responses. We also calculated descriptive statistics and paired t-tests with Bonferroni correction.

\subsubsection{System Usability Scale Scores}

\begin{figure}[h]
  \centering
  \includegraphics[,alt={Bar chart of the mean SUS scores of 20 participants comparing Facilitated English, Natural Deaf Speech, and Touch input methods. Touch scoured highest with a mean of 63.5 and an error rate of 4.657, followed by Facilitated English with a mean of 62.5 and an error rate of 5.059, then Natural Deaf Speech with a mean on 60 and an error rate of 3.559.}]{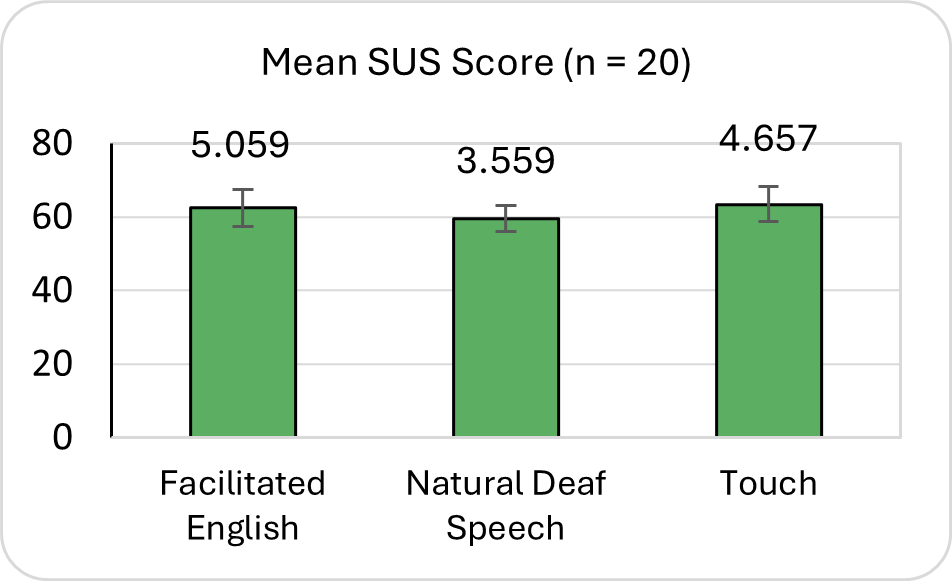}
  \caption{Mean SUS scores for all three conditions.}
  \label{fig:sus2}
  \Description{Bar chart of the mean SUS scores of 20 participants comparing Facilitated English, Natural Deaf Speech, and Touch input methods. Touch scoured highest with a mean of 63.5 and an error rate of 4.657, followed by Facilitated English with a mean of 62.5 and an error rate of 5.059, then Natural Deaf Speech with a mean on 60 and an error rate of 3.559.}
\end{figure}

\begin{figure}[h]
  \centering
  \includegraphics[width=3in,alt={3D scatterplot of the participants’ individual responses for SUS across the conditions of natural deaf speech, facilitated English and LLM-assisted touch. There is significant variability in the two speech-based options with the points spread out in a 2D cloud. The cloud overall appears to be tilted in the Z axis, with better speech SUS also favoring better touch SUS overall.}]{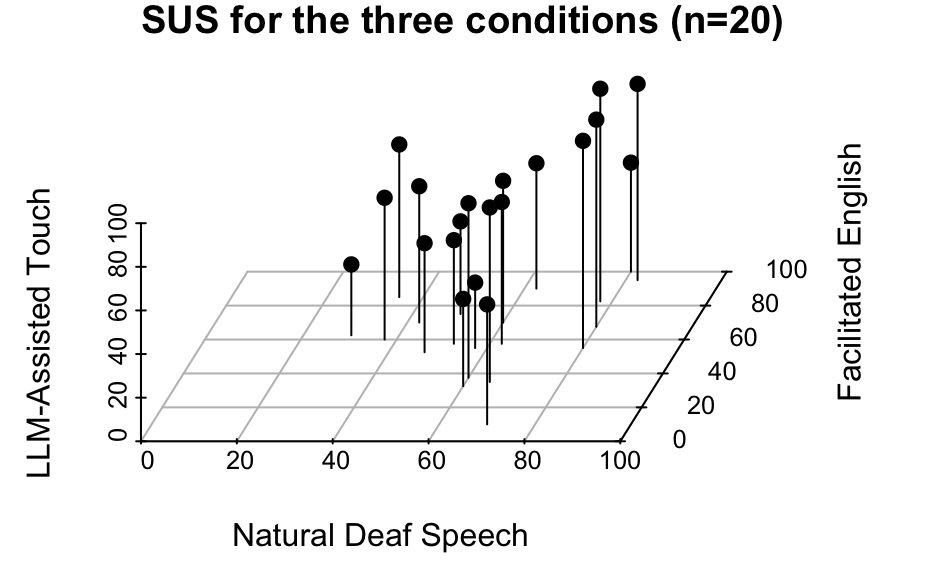}
  \caption{Scatterplot of SUS for all three conditions. The height bars indicate the SUS for the LLM-assisted touch condition, while the x and y coordinates of each point indicate the SUS for the Natural Deaf Speech and Facilitated English conditions for each participant, respectively.}
  \label{fig:susscatter}
  \Description{3D scatterplot of the participants’ individual responses for SUS across the conditions of natural deaf speech, facilitated English and LLM-assisted touch. There is significant variability in the two speech-based options with the points spread out in a 2D cloud. The cloud overall appears to be tilted in the Z axis, with better speech SUS also favoring better touch SUS overall.}
\end{figure}

The SUS scores are shown in Figure~\ref{fig:sus2}. Natural Deaf Speech received a mean SUS score of 59.6 (SD 15.9, SE 3.555), Facilitated English received a score of 62.5 (SD 22.624, SE 5.059), and touch received a 63.5 (SD 20.828, SE 4.657). The paired t-test with Bonferroni correction showed that the differences were not statistically significant. Figure~\ref{fig:susscatter} shows a 3D scatterplot along all three conditions with the results on a per-participant basis. 

\subsubsection{Adjective Scale Scores} 

\begin{figure}[h]
  \centering
  \includegraphics[,alt={Bar chart of the mean adjective score of 20 participants comparing Facilitated English, Natural Deaf Speech, and Touch input methods. Touch scored the highest with a mean of 5.2 and an error rate of 1.348, followed by Natural Deaf English with a mean of 5 and an error rate of 1.451, then Facilitated English with a mean of 4.8 and an error rate of 0.304.}]{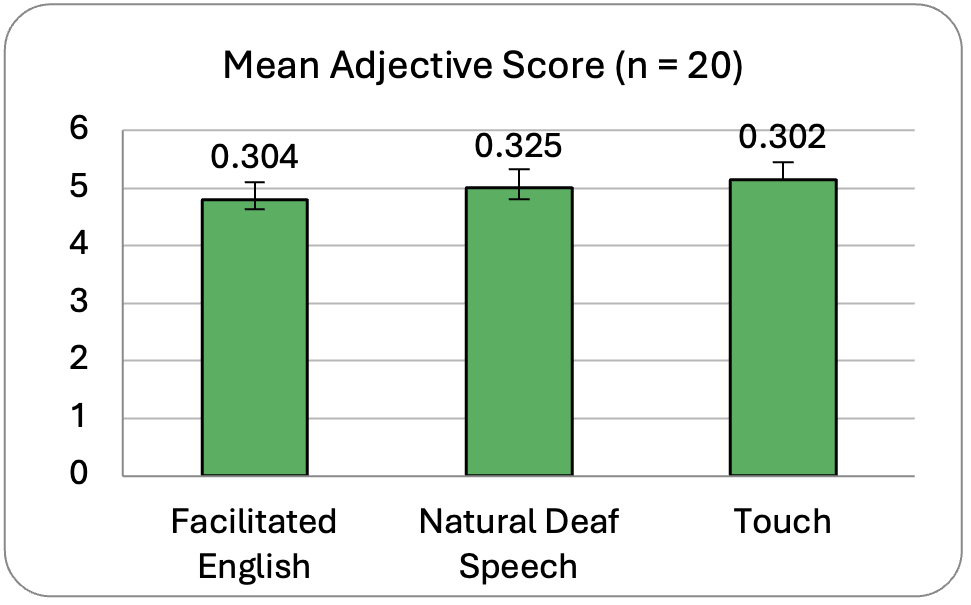}
  \caption{Mean Adjective Scale scores for all three conditions.}
  \label{fig:adj2}
  \Description{Bar chart of the mean adjective score of 20 participants comparing Facilitated English, Natural Deaf Speech, and Touch input methods. Touch scored the highest with a mean of 5.2 and an error rate of 1.348, followed by Natural Deaf English with a mean of 5 and an error rate of 1.451, then Facilitated English with a mean of 4.8 and an error rate of 0.304.}
\end{figure}

The adjective scale scores are shown in Figure~\ref{fig:adj2}. Facilitated English received a mean adjective scale score of 4.8 (SD 1.361, SE 0.304), Natural Deaf Speech received a score of 5 (SD 1.451, SE 0.324), and touch received a score of 5.15 (SD 1.348, SE 0.302). The paired t-test with Bonferroni correction showed that the differences were not statistically significant. 

\subsubsection{Net Promoter Score}

\begin{figure}[h]
  \centering
  \includegraphics[,alt={Bar chart of the NPS based on data from 20 participants comparing Facilitated English, Natural Deaf Speech, and Touch input methods. Facilitated English received a real score of negative 40 and an expected score of negative 30. Natural Deaf Speech received a real score of negative 5 and an expected score of negative 36. Touch received a real score of negative 10 and an expected score of negative 27.}]{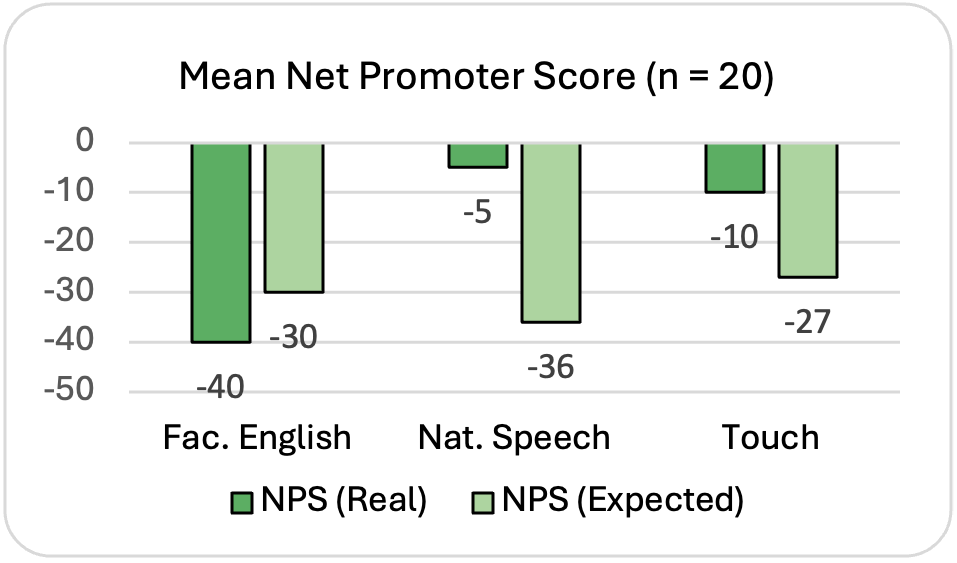}
  \caption{Mean NPS scores for all three conditions.}
  \label{fig:nps2}
  \Description{Bar chart of the NPS based on data from 20 participants comparing Facilitated English, Natural Deaf Speech, and Touch input methods. Facilitated English received a real score of negative 40 and an expected score of negative 30. Natural Deaf Speech received a real score of negative 5 and an expected score of negative 36. Touch received a real score of negative 10 and an expected score of negative 27.}
\end{figure}

Typically, Netpromoter scores can be predicted from SUS scores~\cite{sauro2012predicting}. The predicted NPS scores were -36 for Natural Deaf Speech, -30 for Facilitated English, and -27 for touch.  However, the observed NPS scores were -5 for Natural Deaf Speech, -40 for Facilitated English, and -10 for touch. Both are shown in Figure~\ref{fig:nps2}.

\subsubsection{Word Error Rates for Natural Deaf Speech}

\begin{figure}[h]
  \centering
  \includegraphics[,alt={Bar chart of word error rates with respect to Alexa recognition. 50\% of the participants show no errors, then the rate increases quadratically for the subsequent participants to a high of 30\%; followed by a jump to 100\% for the last two participants.}]{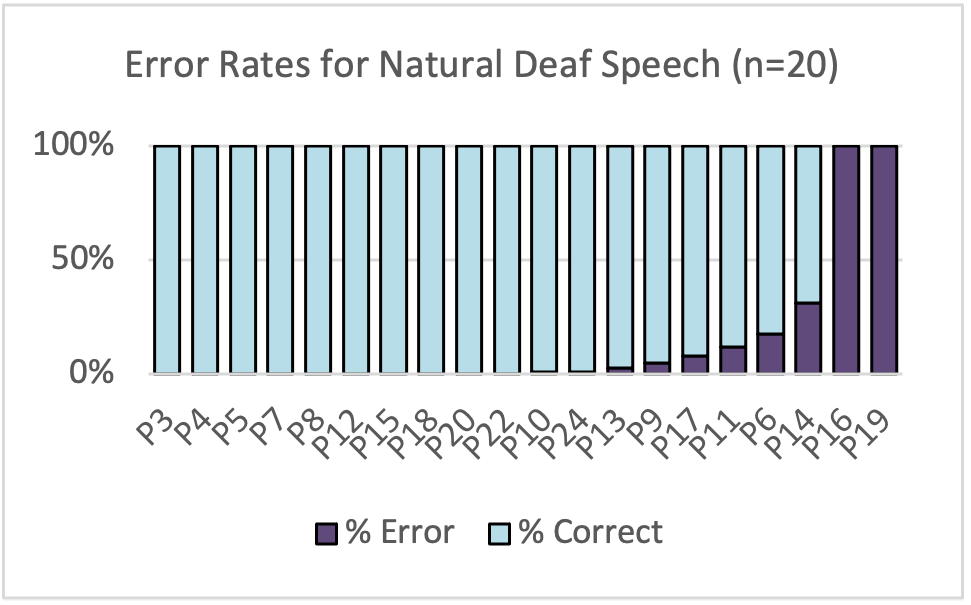}
  \caption{Word Error Rates for the Natural Deaf Speech condition, sorted by participants' Alexa recognition error rates.}
  \label{fig:wer}
  \Description{Bar chart of word error rates with respect to Alexa recognition. 50\% of the participants show no errors, then the rate increases quadratically for the subsequent participants to a high of 30\%; followed by a jump to 100\% for the last two participants.}
\end{figure}

Alexa understood the majority of participants' natural speech with few errors. The WER results are shown in Figure~\ref{fig:wer}. Half of our participants had no Alexa recognition errors at all, which is striking considering that we were initially unsure whether participants would have any success with their natural deaf-accented voices. Two participants were not understood at all by the device with a WER of 100\%, and often failed to even make it past the wake word. The remaining participants had error rates ranging from 0.61\% to 30.91\%. 

We found a strong, statistically significant correlation between WER and SUS scores (r=-0.51, p=.022). There were weaker, non-significant correlations between WER and adjective scales (r=-0.17) and WER and the raw NPS scores (r=-0.4).

\subsubsection{Post-Experiment Survey}

\begin{figure}[h]
  \centering
  \includegraphics[,alt={Pie chart showing that when participants were asked about touch as an alternative input to spoken English, 45\% described it as a very poor alternative, 30\% as a good alternative, 15\% as a poor alternative, and 10\% as a fair alternative.}]{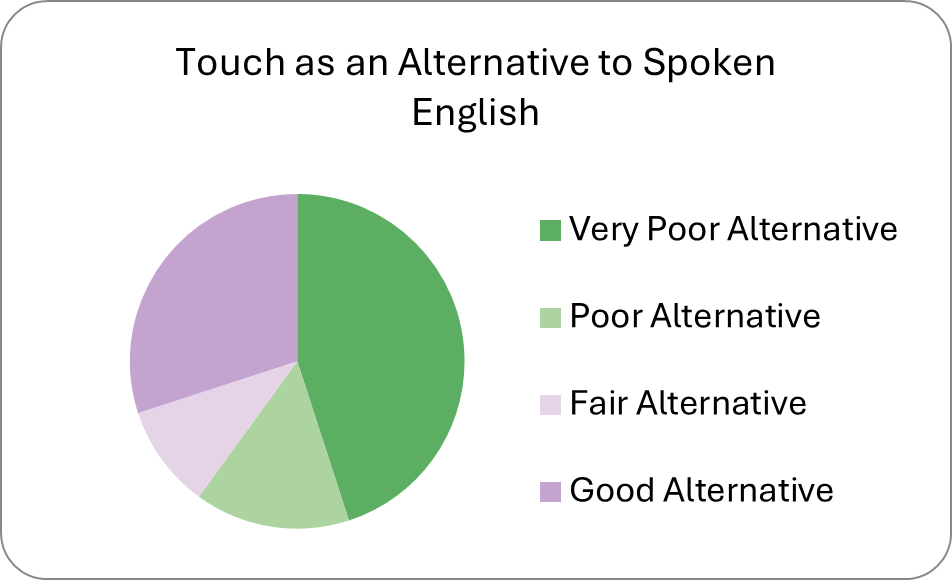}
  \caption{Participants' ratings of touch as an alternative to speech input.}
  \label{fig:engpie2}
  \Description{Pie chart showing that when participants were asked about touch as an alternative input to spoken English, 45\% described it as a very poor alternative, 30\% as a good alternative, 15\% as a poor alternative, and 10\% as a fair alternative.}
\end{figure}

When asked directly about their experiences in the study, 45\% of participants felt that LLM-assisted touch was the easiest to use, 35\% felt Natural Deaf Speech was the easiest, and the remaining 20\% considered Facilitated English easiest.  We also asked about touch as an alternative input method to spoken English, to which more than half responded negatively (Figure~\ref{fig:engpie2}).

\begin{figure}[h]
  \centering
  \includegraphics[,alt={Pie chart showing that when participants were asked about their interest in sign language recognition technology, 45\% described themselves as extremely interested, 30\% as very interested, 15\% as moderately interested, 5\% as slightly interested, and 5\% as not at all interested.}]{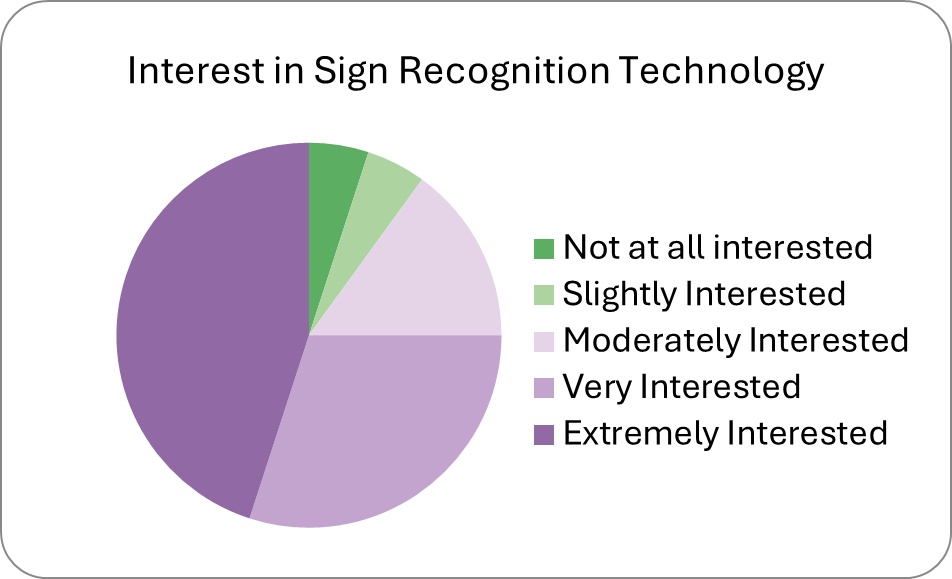}
  \caption{Participants' interest in sign language recognition.}
  \label{fig:recogpie2}
  \Description{Pie chart showing that when participants were asked about their interest in sign language recognition technology, 45\% described themselves as extremely interested, 30\% as very interested, 15\% as moderately interested, 5\% as slightly interested, and 5\% as not at all interested.}
\end{figure}

Because all of our participants had indicated that they used ASL for some of their communication, despite their use of spoken English for the Alexa interaction, we asked them about their interest in using a potential IPA with sign language recognition. Most participants expressed a high degree of interest in such technology (Figure \ref{fig:recogpie2}).

\subsection{Qualitative Results: Interview Themes}

\begin{figure*}[h]
 \centering
  \includegraphics[,alt={The thematic map illustrates the key themes related to input method: voice, input method: touch, and technology adoption. The sub-themes are as follows: (1) Input Method: Voice: (1a) Natural Deaf Speech, (1b) Facilitated English, (1c) Experimental Design, and (1c1) Use of WoZ. (2) Input Method: Touch, (2a) AI, (2a1) Data retention, (2a2) Learning habits, (2b) UI/UX, (2b1) Touch options limited, (2b2) Touch latency, (2b3) Wants improvements. (3) Technology Adoption, (3a) Accessible Alternatives, (3b) Interested in using IPAs more, (3c) Hands-free, (3c1) Signing/ASL, and (3c2) Gestures.}]{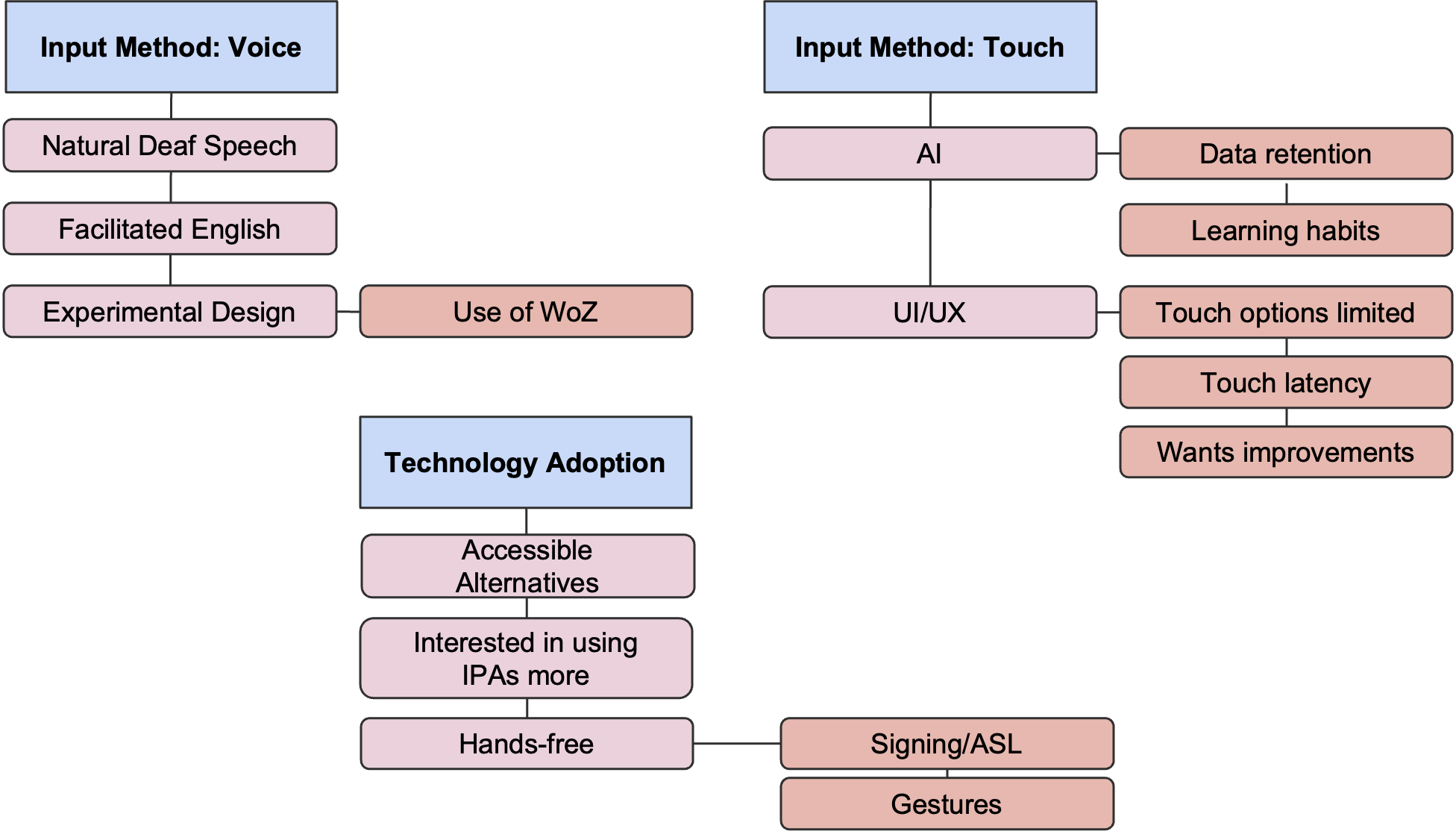}
  \caption{Theme map. Overall, three major theme clusters were identified.}
  \label{fig:themes}
  \Description{The thematic map illustrates the key themes related to input method: voice, input method: touch, and technology adoption. The sub-themes are as follows: (1) Input Method: Voice: (1a) Natural Deaf Speech, (1b) Facilitated English, (1c) Experimental Design, and (1c1) Use of WoZ. (2) Input Method: Touch, (2a) AI, (2a1) Data retention, (2a2) Learning habits, (2b) UI/UX, (2b1) Touch options limited, (2b2) Touch latency, (2b3) Wants improvements. (3) Technology Adoption, (3a) Accessible Alternatives, (3b) Interested in using IPAs more, (3c) Hands-free, (3c1) Signing/ASL, and (3c2) Gestures.}
\end{figure*}

The theme map for the semi-structured post-experiment interviews is shown in Figure~\ref{fig:themes}. We identified three major clusters of themes: Input Method: Voice, Input Method: Touch, and Technology Adoption. These clusters capture participants' experiences with both voice and touch interaction methods, including perceptions of natural vs facilitated speech, feedback on the role of WoZ as the experimental design, concerns related to AI and UI/UX in the touch interface, and the possibility of more accessible, hands-free alternatives such as signing/ASL, and gestures.

\subsubsection{Input Methods: Voice}
\label{sec:voicequal}

Participants were split on their opinions of both voicing conditions.  As previously mentioned, we included Facilitated English as a comparison to the Natural Deaf Speech condition, because we had initially been unsure whether Alexa would succeed with \textbf{Natural Deaf Speech}. Once the study began, it became clear that Alexa could understand some speech from individuals who are deaf.\textit{``When I saw that it [Alexa] understood every command I gave it, I was shocked'' (P17); ``I was impressed with how much the device understood me'' (P6); ``I'm scared to say, tell me the weather, but because it has worked with this Alexa device, I'm going to go home and test with my device and see how much it can understand me... It seems like it's possible that I could have been doing this all along and I just had no idea.'' (P9); ``I've always resisted using any voice activated devices. I'm deaf. I wondered how they could possibly benefit me and didn't care. But, I was impressed with how much the device understood me. So, now I'm curious... maybe I should try it sometime. I should try getting a device'' (P6)}

In contrast, some participants struggled to be understood by Alexa at all. \textit{``I noticed my deaf voice wasn't recognized. I struggled with the second condition [Natural Deaf Speech]. I grew up, raised orally. I worked in the government with other hearing people...they would understand me. So, the problem is that the deaf voice is unrecognizable, and I struggled'' (P16); ``Voicing is hard for me... When I say a word it can be hard to understand'' (P11)}.

While some participants appreciated the accuracy of \textbf{Facilitated English} \textit{``The third one I loved because the facilitated interpreter or something'' (P16)}, others struggled with the delays that it induced. \textit{``I wasn't sure if it was hearing me or not. There's a little bit of a delay. I wasn't sure if I should repeat myself'' (P7)}. \textit{``Yes, it takes more time than if I'm speaking from myself directly to the device'' (P24)}. One participant mentioned that having captions as feedback for their voice could help with the delay. \textit{``[W]ith Facilitated English, it was hard for me to catch everything. Like, I was like, okay, did you get my command? I think it was maybe [a] lag time. [...] I wonder if you could add captions like whatever is said so I can know if Alexa is wrong [...]'' (P14)}. Also, not all deaf speech was understood by the human wizard in this condition. 

Participants had thoughts about the \textbf{experimental design} related to the voice input methods. Despite efforts to distinguish the two voice conditions, several participants were uncertain about how their speech was processed during the study. Some were unsure whether the recognized speech came from their own voice or from the facilitated re-speaking, \textit{``I don't know, it was weird. Right. The first one, which one was that?'' (P3), ``I wondered whether that meant that I was speaking so and so correctly, or did that mean there were some program modifications?'' (P9)} One participant questioned whether the facilitator intervened only when needed or if it was always active, asking, \textit{``If my speech is clear, do they still add it in? How does that understand me better than Alexa?'' (P13)}

A few participants were surprised to learn that the Facilitated English condition had a \textbf{Wizard-of-Oz} design involving a live human interpreter, and initially assumed the system was automated: \textit{``I thought maybe it was a robot.'' (P10)}. We observed mixed responses to the human wizard during the interview with participants, and the same researcher played both the wizard and the interviewer roles. During the interview portion of the study, one participant expressed that the consent form did not clearly describe the role of the facilitator, which reveals a weakness in the experimental design. 

\subsubsection{Input Methods: Touch}
\label{sec:inputtouch}

Most participants expressed a positive attitude towards the LLM-assisted touch interface. \textit{``Touch was really good for me. [...] It was easy to use, and I saw what the options were, so I liked that.'' (P4).} \textit{``Yeah, touch was my favorite for sure.'' (P5).} It was also mentioned as a solution for situations when Alexa didn't understand a person's speaking: \textit{``It was the easiest to use. Easier than using voice, which was a struggle to make sure that I was getting them right.'' (P11).} 

Many participants were comfortable with the concept of \textbf{AI} \textbf{learning their habits} for more accurate predictions. \textit{``I think it's fine. It should make things easier if I touch it over and over again then it should make things easier and be all right.'' (P4); ``I don't really feel like I have privacy. I mean, what I did is data. AI accessing that, I mean, it's just data, right?'' (P9).}

However, some expressed reservations: \textit{``I try to avoid using my personal information and feeding that to AI because, as you said, it learns and it's work and progress as well, it can be a little scary.'' (P7).}

Participants who had reservations wanted more control over \textbf{data retention}, and the option to delete their past history and personal data, but they still saw some value in retaining it. \textit{``I would also like the option to have it forget what it knows about me, because there are some times where I google something and I don't want it to remember that like I ask a weird question or I use incognito search. I want to be able to hide it so that the machine doesn't judge me.'' (P6); ``I don't want to have to wipe everything. Yeah, I want to pick what I delete or wipe. [...] I want a full wipe or like have the option to go through my search history.'' (P20); ``Are [friends] borrowing it or am I gifting it to them? If they're borrowing it, then I'd be fine with just deleting portions of the data [...]'' (P17).}

The LLM approach also resulted in other \textbf{UI/UX} peculiarities that were noticed by participants; one of these was the variability in LLM responses, which caused options to show in an unpredictable order. \textit{``Oh, right, yes, oh, they moved. [...] Yeah, I noticed.'' (P3).} Sometimes the \textbf{touch options were limited}; the ones displayed on the grid did not include the specification users were looking for. \textit{``I'm a little OCD, so that is not something that I will accept if it's 75\%, it's 75\%'' (P10)}. However, a minority of participants were satisfied with selecting the buttons that displayed the closest approximation: \textit{``I saw plenty of options, but eighty-five and eighty, that's a five percent difference. I don't care'' (P6)}.

Several participants commented on the \textbf{touch latency}, with a response time of approximately two seconds. The latency was a direct consequence of querying the LLM APIs. \textit{``Touch felt like it took forever. and I had to really pay attention. I couldn't really like multitask, and then I had to confirm: is that what I want? [...] I have no patience.'' (P6); ``I mean, I still think it's cool. [...] It's definitely slow with the response, the delay time. There's definitely a [slight] lag when accessing it. That's something that you could maybe work on'' (P18)}. One participant mentioned that typing was more efficient: \textit{``It's AI, so I understand, but it was slow [...] I could type once I was told I could type, but using the buttons was not natural.'' (P9)}.

Participants appreciated having a touch interface available; however, several \textbf{wanted improvements}. One participant suggested that it would be helpful to \textit{``make it a button that says input your own words, just to let them know that they can type if the option is not there'' (P9}). Others felt constrained by the number of available choices, explaining that \textit{``if you just have these few button options, you're a little bit stuck'' (P19)}. There was also some confusion about the intended functions of certain buttons, including one participant who asked, \textit{``Remember like the orange one that was Alexa stop? What was that for?'' (P9)}.

Participants also commented on the interface layout and efficiency, describing the design as visually \textit{``a little bit overwhelming''} and saying that it sometimes required \textit{``at least three to four clicks to finally get what I wanted'' (P20)}. Suggestions for improvement included allowing users to type full commands and creating shortcuts for commonly used actions, such as saving frequent commands in \textit{``the heart storage''} (P22) for quicker access. 

\subsubsection{Technology Adoption}

Participants reported that having different \textbf{accessible alternatives} for input options would strongly influence whether they would use an IPA. Some said they might use speech in specific situations, such as driving, but many did not want to rely on voicing for everyday interactions. One participant explained they \textit{``don’t want people ... to depend on [them] and the use of [their] voice'' (P6)}, describing how speech carries social expectations they prefer to avoid. Typing was seen as a helpful alternative, with participants noting that it could be \textit{``more efficient'' (P7)} and that they liked \textit{``being able to use something where I can type'' (P15)}, especially when speech clarity varies, or voice use is tiring.

Participants who could envision accessible input options were \textbf{interested in using IPAs more} in their daily lives. They suggested tasks such as \textit{``ordering food, shopping… unlocking my door ... TV, timers ... more daily functions'' (P6)}, and one participant compared the interaction to \textit{``a friend that responds to you'' (P13)}. Across interviews, participants described that input flexibility would make IPAs more practical, reliable, and relevant for everyday use.

The importance of having a \textbf{hands-free} input method was a heavily mentioned theme throughout interviews. When asked if participants could think of a hands-free input method that did not rely on speech, several participants expressed the desire for \textbf{signing/ASL} recognition in common household IPAs. \textit{``I think it's important. Like, sometimes I'm doing stuff, or I have my hands on my phone, and I'm cooking. I don't want to touch the device, but I want to interact with it. [...] It's already there, so you know it'd be nice for it to be able to understand my signs.'' (P14)}. Another topic related to the concept of hands-free communication for DHH users is emergency situations. \textit{``I would rather it be hands-free. If I fall and I'm hurt, I want to be able to yell, call 911.'' (P16)}.

Participants described hands-free interaction as essential, particularly when hands are occupied or speech is unavailable. They emphasized that signing should be recognized in home devices and highlighted opportunities to support a broad range of users. As one noted, sign-based access \textit{``should be able to benefit any non-verbal groups'' (P9).} \textbf{Gesture-based} control and eye-tracking were also seen as promising pathways, with one participant adding that \textit{``eye tracking technology… Deaf people would really appreciate that'' (P19).}

Many participants were unsure what other options besides ASL and speech could be used for hands-free interaction. One suggested eye tracking and gesture: \textit{``Yeah, like eye tracking voice, signing or like simple gesture systems. That's it.'' (P19),} as well as blinking: \textit{``I think in perception of machine learning, blinking is very obvious.'' (P9)}.

\section{Discussion}
\label{sec:discussion}

On a high level, the quantitative and qualitative findings provide answers to our research questions (Section~\ref{sec:rq}), as follows.

With respect to RQ1, an AI/LLM-assisted touch interface is a potentially effective alternative to spoken English given its non-significant differences in SUS and Adjective Scale scores. The participant comments, however, identified shortcomings in the implementation that are discussed in detail below, and overall there still is a need for supporting spoken English.

With respect to RQ2, there were no significant quantitative differences between Wizard-of-Oz Facilitated English and participants using their Natural Deaf Speech. However, opinions were polarized, with many participants expressing surprise at how well Alexa worked with their own voices, but others getting frustrated. Many participants did not mind the concept of having an `interpreter' mediate their communication, but some did not appreciate the induced time delays due to the interpretation. This, too, points at the importance of supporting  natural Deaf speech natively in IPAs.

Below, we discuss the quantitative and qualitative findings in more detail.

\subsection{Quantitative Results Interpretation}

\subsubsection{System Usability Scale Score Interpretation}

 The SUS scores fell within the acceptable range~\cite{sauro5ways}. Touch got the highest score, followed by Facilitated English and then Natural Deaf Speech, with no significant difference. As the standard deviations indicate, there was a large spread in opinions across participants. Of particular note, from Figure~\ref{fig:susscatter}, is that Facilitated English showed the most extreme spread in usability scores on either end --- some participants thought that it provided superior usability, while on the other end, some participants hated it. There also is no apparent relationship between Natural Deaf Speech and Facilitated English usability scores. Participants liking one spoken English condition was not predictive of them liking or disliking the other one. Our participants were somewhat more likely to rate the LLM-assisted touch condition favorably when they also had rated one of the spoken language conditions highly. 

Since this work is closely related to Tran et al~\cite{tran2024assessment} and we used comparable usability measures, we can directly compare the methods within the two experiments. In our study, LLM Touch scored as SUS of (63.5, SD 20.828, SE 4.657), which can be compared to both the Tap to Alexa and Apps conditions in Tran et al. It shows an increase from Tran et al's Tap to Alexa condition (61.4, SD 19.304, SE 4.025) and a large increase from the Apps condition (56.3, SD 26.670, SE 5.561). We attribute this increase to the addition of the LLM interface. It is within the medium range of marginal usability based off the SUS score. Based on the participants' comments, latency was a major factor in their evaluations, and it is possible that touch would have scored even higher with lower latencies. A relatively simple engineering fix for hiding latencies, with a high potential payoff, would be pre-fetching the possible LLM responses as soon as the user selects an option.

Ranked just below the touch condition in this experiment was the Facilitated English condition, also in this study with a SUS of (62.5, SD 22.624, SE 5.059). This condition showed the widest standard deviation, which aligns with our qualitative results that participants had many different views on the inclusion of the Facilitated English condition, further discussed in the qualitative results section. Natural Deaf Speech from the current experiment (59.6, SD 15.9, SE 3.555) is on the low end of marginal usability based on the SUS score.  While for half of the sessions, Alexa seemed to understand participants very well, there were two participants where the device was not registering participants' natural Deaf voices at all, which shows in their SUS evaluations (see also the discussion of WER below). In contrast, the natural language condition in Tran et al was ASL, with an SUS of (71.6, SD 16.428, SE 3.427). This score is considered acceptable in terms of usability, and substantially higher than both the Facilitated English and Natural Deaf speech scores we observed in this study. In sum, there is major room for improvement in deaf speech interaction, compared to previously studied ASL-based interaction with IPAs.

Also closely related is DeVries et al~\cite{devries2024sign} which measured ASL use with sticky hands as well as apps with Alexa. The mean SUS score for the ASL condition in that publication were very similarly rated to Tran et al's ASL mean SUS score, and also substantially higher than our Facilitated English and Natural Deaf Speech scores. We can also compare our LLM Touch condition to the Apps to Alexa with sticky hands condition from that publication. Here, too, our LLM Touch condition performed better, but it is uncertain to what extent the difference is attributable to having the predictive LLM support, and to what extent having sticky hands interfered with app use in DeVries et al.
 

\subsubsection{Adjective Scale Score Interpretation}

 The adjective scale that we applied asked participants to rate the user-friendliness of the system on a seven point scale with one being worst imaginable to seven being excellent.  Touch received the highest score at 5.15 (SD 1.348, SE 0.302), closely followed by Natural Deaf English at 5 (SD 1.451, SE 0.324), and Facilitated English at 4.8 (SD 1.361, SE 0.304). These scores indicate that all three input methods ranged from “ok” to “good” user-friendliness through the eyes of the participants. Those results show both promise with the baseline “ok” or even “good” met and room for improvement with the scores of “excellent” yet to be received.

 Tran et al did not note Adjective Scale or NPS scores, though DeVries et al~\cite{devries2024sign} did so we can compare to those.  For apps, DeVries reported the Apps condition at 4.7 (SD 1.4, SE 0.255) which is low compared to the touch condition of this experiment, suggesting an improved user experience with the addition of the LLM-supported touch interface in comparison with what is currently offered in terms of input methods with Alexa, such as Apps and Tap to Alexa. DeVries et al also employed an ASL with sticky hands condition  which came out to 5.7 (SD 1.2, SE 0.218), a score higher than any of the three conditions in this experiment. This again indicates that there is more room for improvement in deaf-accented speech interaction with IPAs than for ASL-based interactions from previous studies.

\subsubsection{Net Promoter Score Interpretation}

SUS scores are typically predictive of NPS scores and vice versa~\cite{sauro2012predicting}. However, past related work found a discrepancy between these two for the case of DHH Alexa usability evaluations~\cite{devries2024sign}. To assess whether this discrepancy also exists for DHH evaluations involving spoken English, we compared the expected NPS based on the SUS scores and the observed NPS. Touch NPS was rated slightly more highly than expected, and Natural Deaf Speech was rated much more highly than expected. As with the other two usability measures above, Facilitated English generated the most extreme spread across participants for NPS. The case for improving Natural Deaf Speech support in IPAs is bolstered by the significant gap in the expected NPS for Natural Deaf Speech and the actual score, which was much higher than anticipated.  This aligns with our qualitative results that participants' were excited to use their voices with Alexa (as long as Alexa understands them). It is also possible that this excitement was partially driven by the fact that some participants were surprised at how well Alexa understood them.

\subsubsection{WER Interpretation}

Alexa's recognition accuracy on deaf-accented speech was surprisingly high, given that half of our participants exhibited no recognition errors, and another quarter of participants exhibited error rates under 10\%. This is likely good enough to support real-world interaction scenarios, which is also backed up by the interview results where some participants stated that they were now more motivated to try Alexa at home. However, this result should be interpreted cautiously with respect to the general state of the art in deaf-accented ASR. A command-and-control interface uses well-defined patterns for interaction that, more likely than not, are at play here when Alexa interprets commands. This is evidenced in part by the fact that, as mentioned in Section~\ref{sec:measures}, Alexa ignores filler words and false starts. Thus, the constrained environments under which Alexa operates, very likely contributed to a substantial reduction in recognition errors, simply because of the limited number of options that make sense for a sequence of words in a command. A general-purpose ASR transcription system, as opposed to an IPA, has to contend with a much greater space of options, and continues to struggle with deaf-accented speech. Nevertheless, this result is encouraging for the state of DHH access to IPAs, and strongly suggests that it is worthwhile to invest additional resources into improving ASR in command-and-control contexts. It also remains to be seen how well more advanced, less-constrained chatbot-like IPAs will perform, such as the recently announced Alexa+.

\subsection{Qualitative Results Interpretation}

Many of the qualitative findings center around delays and latencies. To put these in context, here are the average latencies for each condition: (1) Natural Deaf Speech: 2 seconds from command to Alexa response; (2) Facilitated English: 3 seconds for listening and re-speaking plus 2 seconds to Alexa response for a total of 5 seconds; (3) LLM Touch: 6--8 seconds to build the command plus 2 seconds for Alexa response, for a total of 8--10 seconds. Note that while these 8--10 seconds may appear high, this is likely less than the average time it would take to navigate the tasks via smart home apps or Tap-to-Alexa, as done in prior literature~\cite{tran2024assessment}.

\subsubsection{Interview Themes Interpretation}

Our findings show that DHH users have mixed experiences with current speech-based IPA systems. While some participants were surprised and pleased when Alexa accurately recognized their speech, others experienced repeated failures and frustration. These mixed outcomes underscore the ongoing need for IPA systems to more effectively support diverse Deaf speech characteristics, rather than relying on users to adapt to hearing-centric models.

Facilitated English provided higher recognition accuracy for participants who struggled with direct interaction. However, delays from human re-speaking introduced uncertainty and reduced the feeling of natural interaction. Because several participants did not clearly understand the facilitator’s role until after the study, this also highlights the importance of transparency when using Wizard-of-Oz methods. However, because in many types of studies, the wizard is essential to a make-believe scenario of technical capabilities --- here, the capability to understand and re-speak deaf-accented speech ---, the disclosure must be managed carefully. Otherwise, there is a risk of participants interacting with the technology differently when they know that a human is in the loop.
 
With respect to touch, our findings show that participants' attitudes are positive toward touch interfaces, and that such interfaces are needed. However, some participants also felt discomfort or overwhelmed during the course of the study. Most participants commented on the delays and the limited range of options they had available; there was no freedom to customize the touch UI buttons. It was for this reason that some participants indicated a preference for the context-aware on-screen keyboard that we provided as part of the LLM touch interface, as a seemingly faster alternative. There clearly is a need for UI and UX refinements to our current implementation.

Participants emphasized that accessibility and adoption are tied to having control over communication. Many reported that voice is not always a comfortable or preferred way to communicate, even when it technically works. Typing was valued as more reliable, and hands-free options such as signing were described as essential for everyday contexts like cooking or emergency situations. These perspectives suggest that future IPAs should support flexible, multimodal interaction rather than relying only on speech.

\subsubsection{Implications for IPA Design}

Notwithstanding the need for multimodal interaction options, the participant experiences and comments show that having functional and reliable natural language options is important. Despite rating the usability of touch comparably to speech on average, the majority of participants stated that touch was a poor alternative to speech. This suggests that in the medium to long term there may be no substitute for making significant improvements to the ability of IPAs to understand deaf-accented and other types of dysarthric speech. 

Overall, how deaf and other dysarthric speech is tackled appears to be important, as evidenced in the large spread of ratings for Facilitated English. While some participants loved it for its accuracy, the delay induced by re-speaking causes significant usability problems. This suggests that rather than relying on third-party re-speaking solutions that translate dysarthric speech into something intelligible to IPAs, the only real solution will be for an IPA to understand deaf or dysarthric speech natively.

However, beyond speech, there is still a need for robust multimodal interaction options. LLM-based touch can be an appropriate fallback if the identified problems with LLM latency are addressed and the context-dependent choices of the available touch options are improved. Beyond touch, designers also need to consider other options to interact with IPAs. Based on the participants' comments, the range of options may be highly situational; such as hands-free options for cooking, or typing if the user's confidence in their speech is low.

\section{Limitations and Future Work}

\subsection{Limitations}

One limitation is a result of the Wizard of Oz methodology for the facilitated spoken English condition. Having a researcher behind the scenes acting as an 'interpreter' represents a best-case scenario for re-speaking deaf-accented speech, for two reasons. First, there is no training involved, and second, it is likely that our trained human performed better at understanding deaf speech than available technology would at this stage. This may have presented the Facilitated English condition in a better light than it would be in the real world, despite the participants frustrations with the latency (which, however, also would show in real-world systems). An additional limit of using a human wizard is the introduction of human error. 


Our participant sampling focused on people who use spoken English. However, every participant also indicated that they use ASL for their everyday communication. Although some of our participants conducted all their interactions in spoken English, including the semi-structured interviews, this means that our sampling of deaf-accented speech cannot be representative of the entire spectrum of DHH people and more research is needed. 

Another limitation was the unpredictability of the LLM within the touch interface. In an effort to narrow the range of API outputs, we provided extensive contextual information. This included providing a list of example tasks that were similar (but, importantly, not the same) as the tasks  participants were instructed to perform. 

The downside of this approach is that the usability for this model would likely vary for a real-life scenario because spontaneous tasks cannot be primed in the GPT model. However, the API should learn over time from user history, so it is likely that slight modifications to the code paired with repeated use would counteract the restrictions we implemented for our limited domain application.

\subsection{Future Work}

Given the success of using LLMs for touch input, this technology would benefit from future research to further its development. Researchers might consider evaluating the usability of APIs with different creative temperatures or change the way in which tasks are queried and presented. It may be insightful to enable the interface for more complex Alexa interactions, such as maintaining contextually aware conversations. For example, if a user tells Alexa to play music, then Alexa follows up and asks, “What type of music would you like to listen to?” A more developed interface could accommodate the complexities of a two-sided conversation between Alexa and the user.

The focus of further studies should reflect participants' desire for hands-free IPA interaction. In particular, researchers may consider integrating an LLM interface controlled through sign and Deaf speech. Another avenue for future research is spatial interfaces, such as an interactive projection.

One potential way to improve responsiveness would be to fetch appropriate LLM API responses to every option that the user can select in advance; however, we did not implement that improvement in this study.

Another important avenue for future work is testing re-speaking approaches with commercially available technology. It will be important to isolate the effect of the time spent on training the systems from the interaction with the IPAs in order to arrive at meaningful usability results.

\section{Conclusion}


While IPAs, like Amazon Alexa, already offer voice-controlled functionality, their ASR systems still struggle to understand atypical speech like that of some DHH individuals who use their voice. The results of this study indicate that current ASR systems in IPAs are capable of understanding some DHH individuals' speech. We recognize it as a viable option, with opportunities for improvement. When we set out to do this study, we had expected that a re-speaking solution with ASR, specifically tailored to each DHH individual, would be a necessary interim step in providing access to IPAs. However, the results and participant comments from the facilitated English condition with the Wizard-of-Oz human re-speaking now make this approach appear much less desirable. Given that a trained human is probably more capable of understanding deaf-accented speech than even the best automated re-speaking technology, the findings in this study are probably near the ceiling of what is possible with re-speaking. The re-speaking delays, and possible elements of confusion for DHH users who pick up on the re-spoken audio, appear to outweigh the potential benefits in many instances. 

 Our findings from this study complement prior work that has explored ASL as an alternate natural language input for smart devices and IPA interfaces. They underscore the need for flexible, multimodal IPA input options, such as natural language options (ASL and Natural Deaf Speech), as well as a touch-based, LLM-supported interface, along with other situational options, such as gesturing or typing. Input options must be manageable by a wide range of users. Participants not only wanted the device to recognize their commands, but also to be customizable, responsive, and aware of context. Furthermore, participants recognized the importance of hands-free interaction. To support this goal, many expressed an interest in having natural language input options to interact with IPAs in the forms of both ASL and Natural Deaf Speech, showing that these two options are not mutually exclusive even within the same person.
 
  With respect to natural language options, it is important to recognize the limitations of current technology. Even though Natural Deaf Speech input performed surprisingly well on our participants, it still remains inaccessible to a substantial fraction of DHH people who exhibit atypical speech. Similarly, ASL-based interaction has been tested only in Wizard-of-Oz studies in related work, but not yet been deployed in the wild. This means that natural language options still require further research and development before they can become part of universal design practices in IPAs. This study also validates the prospects of LLM-generated task suggestions as an alternative to natural language input for interactions with IPAs. The touch-based LLM interface tested in this study, however, still poses significant latency issues.  Overall, the key takeaway from this paper is that both Natural Deaf Speech and LLM-assisted touch interfaces show substantial promise for future research and implementation, and should be pursued toward the goal of making IPAs accessible.




\begin{acks}
The contents of this paper  were developed under a grant from the \grantsponsor{NIDILRR}{National Institute on Disability, Independent Living, and Rehabilitation Research}{https://doi.org/10.13039/100009157} (NIDILRR grant number \grantnum{NIDILRR}{90REGE0013}). NIDILRR is a Center within the Administration for Community Living (ACL), Department of Health and Human Services (HHS). The contents of this paper  do not necessarily represent the policy of NIDILRR, ACL, HHS, and you should not assume endorsement by the Federal Government. Additional support has been provided by the \grantsponsor{NSF}{National Science Foundation}{https://doi.org/10.13039/100000001} under Awards No. \grantnum{NSF}{2150429}, \grantnum{NSF}{2348221} and \grantnum{NSF}{2447704}.
\end{acks}

\bibliographystyle{ACM-Reference-Format}
\bibliography{bibliography}

\appendix
\section{Interview Guide}
\label{app:guide}

This appendix contains the interview guide.

\subsection{Guide for semi-structured interview with Alexa participants}

Before we do the interview, participants should attempt all three conditions and task lists.

This interview should last no longer than 30 minutes.

The interview will be face to face, but recorded through a Zoom meeting. At the beginning, make clear that the Zoom recording will be used \textbf{only} to analyze the information from the interview, and the video recording of Zoom will not be shared with anyone.

\subsection{Things we would like to learn}

\begin{enumerate}
    \item What works well with the interfaces and what doesn’t?
    \item What are the main things we should improve in the touch interface?
    \item How important is hands-free interaction, and what are the preferred ways to do this?
    \item What else would the participant like to tell us about interacting with Alexa?
\end{enumerate}

\subsection{Questions}

\subsubsection{Category: What works well with the interfaces and what doesn’t?}

\begin{enumerate}
    \item You’ve had a chance to interact with Alexa in three different ways, after a training session with issuing voice commands. What thoughts come to your mind first about this experience?

    \item Probing/follow-up if needed:
    \begin{enumerate}
        \item Anything you especially like about interacting with Alexa in deaf-accented speech? Anything you especially dislike?
        \item Anything you especially like about interacting with Alexa in facilitated speech? Anything you especially dislike?
        \item Anything you especially like about interacting with Alexa with the touch method? Anything you especially dislike?

    \end{enumerate}
    \item If they explain why they like a feature, no follow-up needed. Else, for both features, ask follow-up questions: “You mentioned that X. Why did you like or dislike this?”
\end{enumerate}

\subsubsection{Category: What are the main things we should improve in the touch interface?}

\begin{enumerate}
    \item Let’s talk a bit more about the touch method now. You might have noticed that it changes around a bit every time you use it. That is because it uses artificial intelligence to learn the options. It also remembers what you’ve done with every command and tries to learn your preferences based on that. How do you feel about this idea?

    \item Probing/follow-up:
    \begin{enumerate}
        \item How comfortable do you feel with Alexa learning your habits?
        \item Do you need an option to tell Alexa to forget what it knows about you, even if this means that the touch method will become less convenient to use?

    \end{enumerate}
    \item The touch method might sometimes not show the options that you are looking for. If that happens, how should we improve it?

    \item Probing/follow-up:
    \begin{enumerate}
        \item What do you think about typing on the touchscreen when an option doesn’t show?
        \item What do you think about picking an option that feels “close enough” to what you intended. For example, set the light to 80\% instead of 75\%?

    \end{enumerate}
    \item How else could we improve the touch method?
\end{enumerate}

\subsubsection{Category: If Deaf speech to Alexa is not available, how important is hands-free interaction, and what are the preferred ways to do this?}

\begin{enumerate}
    \item You navigated two ways through this interview of how to interact with Alexa using speech: natural deaf-accented speech, and using a live interpreter behind the scenes. We have been exploring how people use Deaf speech with Alexa. That information is important to companies like Amazon, so that they can build Deaf speech understanding into Alexa in the future. We can’t predict how fast the technology will develop- It could take 2 years, 5. Or more. We don’t know. One nice thing about speaking to Alexa is that you don’t have to touch anything. You can do so even when your hands are dirty while you are working with food in the kitchen. Or you could sign to Siri while you are driving; for example, to find the nearest gas station.

    \item How important is it to you to have hands-free options?

    \item Probing/follow-up:
Are there other ways you can imagine interacting with Alexa without speaking, and without touching a screen?

\end{enumerate}

\subsubsection{What else would the participant like to tell us about interacting with Alexa?}

\begin{enumerate}
\item Is there anything else that you would like us to know about making Alexa better for oral deaf people?
\end{enumerate}

\section{Task Lists}
\label{app:tasklist}

Below are the task lists that we used. Some specific location names have been replaced by placeholders, so as not to break the anonymity of this submission.

Task List 1:
\begin{enumerate}
\item Turn on the lights
\item Dim the lights 50%
\item Show my calendar today
\item Check the weather in my current location
\item Set the light color to blue
\item Show the news
\item Turn off the lights 
\item Set a timer for 2 minutes
\item Show directions to "<name of store>"
\item Show me a recipe for banana bread
\end{enumerate}

Task List 2:
\begin{enumerate}
    \item Show me a recipe for lasagna
\item Turn on the lights
\item Set the light color to red
\item Show the news
\item Show my calendar for tomorrow
\item Dim the lights 25%
\item Show directions to "<name of>" Station 
\item Set a timer for 5 minutes
\item Check the weather in "<name of city>"
\item Turn off the lights
\end{enumerate}

Task List 3:
\begin{enumerate}
    \item Show my calendar for this weekend
\item Check the weather in "<name of city>" 
\item Turn on the lights
\item Set the light color to green
\item Show directions to "<name of>" Station
\item Dim the lights 75%
\item Set a timer for 1 minute
\item Turn off the lights
\item Show me a recipe for spaghetti carbonara
 \item Show the news
\end{enumerate}

\section{LLM Prompts}
\label{app:llmprompts}

This appendix contains the prompts used for the generation of the Alexa command, starting with the initial action verb and the list of options for the chosen action verbs. Additional prompts were used if further follow-up questions to the users were required, along with the response options to the follow-up questions. The final prompt assembled the full Alexa command to issue. The code using these prompts can be accessed at \url{https://github.com/Gallaudet-University/chi2026-ipa}.

\subsection{Initial Action Verb}
Upon starting a new command, the GPT prompt for generating the initial action verbs is described below. The prompt was kick-started with a list of initial ideas for commands, as follows.

\begin{verbatim}
task_ideas = [
    "Turn on the lights",
    "Set the light color to blue",
    "Dim the lights 50%",
    "Turn off the lights",
    "Show my calendar for tomorrow",
    "Show my calendar for next week",
    "Show the weather forecast tomorrow",
    "Show the weather in Chicago",
    "Show directions to Starbucks",
    "Show directions to the farmers market",
    "Show the latest sports news",
    "Set a timer for 1 minute",
    "Set a timer for 3 minutes",
    "Set a reminder to call a friend at 4pm",
    "Show me a recipe for chocolate cake",
    "Play my favorite playlist",
    "Set an alarm for 7am",
    "Send a message to John",
    "Find the nearest coffee shop",
    "Read my latest emails"
]
\end{verbatim}

The actual prompt is shown below. Note how the prompt includes the command history and the smart home environment that the tasks are carried out in.

\begin{sloppypar}
\texttt{"Generate a list (no numbers, no dashes, just the items) of 8 action verbs that are ideal for
starting commands in a voice-controlled assistant interface. These verbs should be straightforward, commonly understood, and versatile enough to apply across various tasks including managing home automation, scheduling events, querying information, and controlling media playback. Provide verbs that ensure clear and concise commands suitable for quick voice interactions. No quotations around the verbs. Some examples of commands are in \textbf{\{task\_ideas\}}. Include a verb to show information as well, such as viewing a calendar or a to-do list. Consider the items in this smart home environment, given in \textbf{\{home\_environment\}} and the user's historical data, which is given in \textbf{\{commands\_so\_far\}} (will be empty to begin), to provide relevant predictions for command verbs. For example, if a user has turned on the lights, a logical next command is to set scenes, turn them off or change brightness. No explanations needed."}
\end{sloppypar}

\subsection{Action Option Prompts}
The follow-up prompt for generating detailed options for the action verb selected is below. Like the prompt for the initial action verb, it considers the user's past history.

\begin{sloppypar}
\texttt{"Generate a list (no numbers, no dashes, just the items) of 8 basic, general Amazon Alexa commands that start with the action verb \textbf{'\{action\_verb\}'} in a smart home context. No quotations. These tasks should be similar to those given in \textbf{\{task\_ideas\}}.
Do not specify details such as rooms (i.e. just 'lights', not 'kitchen lights'). Consider the items in this smart home environment, given in \textbf{\{home\_environment\}}, and do not offer tasks that involve smart home objects or technology besides that list. Also consider the user's historical data, which is given in \textbf{\{commands\_so\_far\}} (will be empty to begin), to provide relevant predictions for commands. For example, if a user has turned on the lights, a logical next command is to set scenes, turn them off or change brightness. Do not include the details of the command (do not specify color, time, date, etc). For example, say 'Set an alarm for' and do not include the time. No explanations needed."}
\end{sloppypar}

\subsection{Assessing Need for Task Detail Prompt}
Some task types require additional detail. The prompt to elicit whether a follow-up question on details was needed was as follows.

\begin{sloppypar}
\texttt{
"You are Amazon Alexa. The user has chosen the task: \textbf{'\{task\}'}. Does this task require more information from the user? Some examples of tasks that do not need more information are 'Turn on/off the lights', 'Turn on/off the TV', etc. Some examples of tasks that do need more information are 'Schedule an alarm', 'Show my calendar', 'Set the temperature to'. Basically, check if the sentence is complete or not. If the task does need more information, respond exactly 'Yes', with no quotes and no other explanation. If it does not need more information, respond 'No'"
}
\end{sloppypar}

\subsection{Additional Task Detail Prompt}
If additional details were needed, as per the LLM response to the previous prompt, the follow-up question was generated from this prompt.

\begin{sloppypar}
\texttt{
"You are Amazon Alexa. The user has chosen the task: \textbf{'\{task\}'}. Generate one most important follow-up question that might be needed to complete this task effectively through an Amazon Alexa command. Consider various settings and preferences that might affect the execution of the task. Also consider the user's historical data, given in \textbf{\{commands\_so\_far\}} (but do not reveal this knowledge to the user). The follow-up question must eliminate ambiguity and be specific to the task at hand. For example, if the task is 'Set the lights to a color,' the model should ask 'What color?'. If the task is 'Show a recipe,' ask 'What specific food would you like a recipe for?' If the question relates to calendars, ask 'What day?'. Don't ask about room or location, assume the user wants tasks within the current unspecified room. No explanation or other text needed besides the question. Do not put quotation marks around the question."
}
\end{sloppypar}

\subsection{Follow-Up Question Response Options Prompt}
The response options to follow-up questions were generated from this prompt.

\begin{sloppypar}
\texttt{
"Generate a list (no numbers, no dashes, just the items) of 8 most common expected responses to the question \textbf{\{question\}}. If (and only if) the question relates to directions, consider popular locations within a 3 mile radius of \textbf{\{location\}}. If (and only if) the question relates to weather, suggest popular locations, including a current location. If (and only if) the question relates to recipes, suggest popular recipes. If (and only if) the question relates to date, suggest options like 'today,' 'tomorrow,' etc. If (and only if) the question relates to light brightness, suggest percentages. If (and only if) the question relates to timers, suggest small intervals of time (seconds or minutes). No quotations or explanations needed."
}
\end{sloppypar}

\subsection{Final Command Generation Prompt}
The final command for Alexa was generated with the following prompt.

\begin{sloppypar}
\texttt{
"The user has chosen the task: \textbf{'\{task\}'}. In response to the question, \textbf{'\{q\}'}, the user has responded \textbf{'\{ans\}'}. Formulate a complete, concise command for an Amazon Echo Show based on this information. Do not include the wake word."
}  
\end{sloppypar}

\end{document}